\documentclass[useAMS,usenatbib]{mn2e}
\usepackage{graphicx}
\usepackage{natbib}
\usepackage{lscape}
\usepackage{longtable}
\usepackage{subfig}
\usepackage{amssymb,amsmath}
\usepackage[T1]{fontenc}
\usepackage{hyperref}
\usepackage{amsmath}
\usepackage{breakurl}
\usepackage{times}
\usepackage{amssymb}

\newcommand{\wise}{\textit{WISE}}
\newcommand{\spitzer}{\textit{Spitzer}}
\newcommand{\planck}{\textit{Planck}}

\newcommand\ion[2]{#1$\;${\small\rmfamily\@Roman{#2}}\relax}%

\def\lsim{\lower0.3em\hbox{$\,\buildrel <\over\sim\,$}}
\def\gsim{\lower0.3em\hbox{$\,\buildrel >\over\sim\,$}}

\title[Quasar-CMB Lensing Cross-Correlations]{Weighing obscured and unobscured quasar hosts with the CMB}
\author[DiPompeo et al.]{M.A. DiPompeo$^1$, A.D. Myers$^1$, R.C. Hickox$^2$, J.E. Geach$^3$, G. Holder$^4$, K.N. Hainline$^2$, \newauthor S.W. Hall$^1$\\
$^1$ University of Wyoming, Dept. of Physics and Astronomy 3905, 1000 E. University, Laramie, WY 82071, USA \\
$^2$ Dartmouth College, Dept. of Physics and Astronomy, 6127 Wilder Laboratory, Hanover, NH 03755, USA  \\
$^3$ Centre for Astrophysics Research, Science \& Technology Research Institute, University of Hertfordshire, Hatfield, AL10 9AB, UK \\
$^4$ McGill University, Department of Physics, 3600 rue University, Montreal, QC, H3A 2T8}

\begin{document}
\date{Accepted 2014 October 31; Received 2014 October 30; in original form 2014 September 18}

\pagerange{\pageref{firstpage}--\pageref{lastpage}} \pubyear{2014}

\maketitle

\label{firstpage}

\begin{abstract}
We cross-correlate a cosmic microwave background (CMB) lensing map with the projected space densities of quasars to measure the bias and halo masses of a quasar sample split into obscured and unobscured populations, the first application of this method to distinct quasar subclasses.  Several recent studies of the angular clustering of obscured quasars have shown that these objects likely reside in higher-mass halos compared to their unobscured counterparts.  This has important implications for models of the structure and geometry of quasars, their role in growing supermassive black holes, and mutual quasar/host galaxy evolution.  However, the magnitude and significance of this difference has varied from study to study.  Using data from \planck, \wise, and SDSS, we follow up on these results using the independent method of CMB lensing cross-correlations. The region and sample are identical to that used for recent angular clustering measurements, allowing for a direct comparison of the CMB-lensing and angular clustering methods.  At $z \sim 1$, we find that the bias of obscured quasars is $b_q = 2.57 \pm 0.24$, while that of unobscured quasars is $b_q = 1.89 \pm 0.19$.  This corresponds to halo masses of $\log (M_h / M_{\odot} h^{-1} ) = 13.24_{-0.15}^{+0.14}$ (obscured) and $\log (M_h / M_{\odot} h^{-1} ) = 12.71_{-0.13}^{+0.15}$ (unobscured).  These results agree well with with those from angular clustering (well within $1\sigma$), and confirm that obscured quasars reside in host halos $\sim$3 times as massive as halos hosting unobscured quasars.  This implies that quasars spend a significant portion of their lifetime in an obscured state, possibly more than one half of the entire active phase.
\end{abstract}


\section{INTRODUCTION}
Photons from the cosmic microwave background (CMB) have traveled from the surface of last scattering at $z \approx 1100$, been deflected by the gravitational potentials of massive structures, and mapped over the whole sky by satellites such as the \textit{Cosmic Background Explorer} \citep[\textit{COBE;}][]{1992ApJ...397..420B}, the \textit{Wilkinson Microwave Anisotropy Probe} \citep[\textit{WMAP;}][]{2013ApJS..208...20B}, and most recently the \planck\ mission \citep{2013arXiv1303.5077P} .  Observationally, this gravitational lensing by large scale structure smooths the CMB temperature power spectrum and introduces correlations between intrinsically independent Fourier modes.  

With sufficient resolution and sensitivity, detection of this non-Gaussian signature is now possible \citep[e.g.][]{2011PhRvL.107b1301D, 2012ApJ...756..142V, 2013arXiv1303.5077P}.  This allows a measurement of the projected mass along a given line-of-sight via various estimators that separate the lensing signature from the intrinsic power spectrum \citep{1999PhRvL..82.2636S, 2001ApJ...557L..79H}.  Because the weight of the lensing kernel peaks at $z \sim$1-2, the use of CMB lensing maps has tremendous power for studying the growth of supermassive black holes during periods of intense accretion, as quasar activity also peaks near these redshifts \citep[e.g.][]{2004MNRAS.349.1397C, 2005MNRAS.360..839R, 2006AJ....131.1203F}.

Quasars are of extensive interest because their properties shed light on how supermassive black holes grow, and on the interplay between nuclear activity and host galaxy properties.  Additionally, their extreme luminosities make them ideal cosmological probes.  However, the \textit{complete} quasar population is not particularly well studied because a significant fraction of quasar activity occurs behind large columns of gas and dust \citep[e.g.][]{1989A&A...224L..21S, 1995A&A...296....1C, Lacy:2013ws}.  Large optical surveys efficiently identify ``unobscured'' or ``type 1'' quasars \citep[e.g.\ the Sloan Digital Sky Survey (SDSS);][]{2000AJ....120.1579Y}, while ``obscured'' or ``type 2'' quasars\footnote[1]{Note that in this paper, the terms type 1 and type 2 do not refer to spectral classifications using broad and narrow emission lines, but classifications based on optical-to-infrared colors, as described in section 2.1} largely fall below the flux limits of these surveys \citep[e.g.][]{2008AJ....136.2373R, 2008AJ....136.1607Z}.

Various models for the cause of obscuration in luminous quasars have been proposed.  One example is pure orientation, where type 1 and type 2 quasars are intrinsically the same but appear different depending on our viewing angle relative to the accretion disk.  This is similar to what is seen in low-luminosity, low-redshift Seyfert galaxies \citep[e.g.][]{1993ARA&A..31..473A}.  Alternatively, larger, galaxy-scale high covering fraction material  \citep{2012ApJ...755....5G} could cause obscuration in some evolutionary phase.  The latter scenario is suggested by models of black hole and galaxy co-evolution \citep[e.g.][]{1988ApJ...325...74S, 2008ApJS..175..356H, 2009MNRAS.394.1109C, 2010MNRAS.405L...1B}.

Studying the host dark matter (DM) halos of type 1 and type 2 quasars can shed light on the nature of obscuration, as simple unified models (e.g. the orientation model) predict no difference in DM halo mass.  For statistically significant results, large numbers of objects are needed --- these have been available for type 1 objects for many years \citep[e.g.][]{2004ApJS..155..257R, 2009ApJS..180...67R}, but only recently have the sample sizes of type 2 objects become significant.  Applying techniques from combined infrared (IR), X-ray and radio surveys \citep{2004ApJS..154..166L, Stern:2005p2563} to data from the \textit{Wide-field Infrared Survey Explorer} \citep[\wise;][]{2010AJ....140.1868W} has led to a dramatic increase in the number of obscured quasars available for systematic statistical study \citep[e.g.][]{2012ApJ...753...30S, 2012MNRAS.426.3271M,  2013ApJ...772...26A, 2013AJ....145...55Y, 2013MNRAS.434..941M}.  While IR data alone allow identification of mixed type 1/2 samples \citep[e.g.][]{2013ApJ...776L..41G}, combining IR and optical data provides a means of separating obscured and unobscured populations \citep{2007ApJ...671.1365H}.

Spatial clustering measurements are a powerful way to measure the typical DM halo mass for a population of objects.  In cosmological models where the universe is dominated by DM, the masses of halos in which galaxies and quasars are embedded drive their clustering properties.  By combining clustering measurements with models for how DM halos collapse at different thresholds \cite[e.g.][]{2001MNRAS.323....1S, 2005ApJ...631...41T, 2010ApJ...719...88T}, halo masses can be derived from the quasar bias ($b_q$), which relates the halo masses of quasars to the underlying DM halo distribution.  

The clustering of type 1 objects has been studied extensively, and multiple studies have found that these objects reside in typical halos of mass $\sim 3 \times 10^{12}h^{-1} \ M_{\odot}$ at a wide range of redshifts \citep[$0 < z < 5$; e.g.][]{2004MNRAS.355.1010P, 2005MNRAS.356..415C, 2007ApJ...654..115C, 2007ApJ...658...85M, 2008MNRAS.383..565D, 2009MNRAS.397.1862P, 2009ApJ...697.1634R, 2010ApJ...713..558K, 2013ApJ...778...98S}.  

Recently, clustering measurements have been made for obscured quasars for the first time.  The first of these studies, \citet{2011ApJ...731..117H}, leveraged the power of a cross-correlation with a large galaxy sample to measure the clustering of \spitzer-selected quasars in the Bo\"{o}tes field. The quasar sample was small (in the hundreds), but they found that type 2 quasars cluster \textit{at least} as strongly as type 1s, and possibly reside in more massive halos.  \citet{2014ApJ...789...44D} used \wise\ to build a much larger sample (in the hundreds of thousands) overlapping SDSS and measured the angular autocorrelation of type 1 and type 2 objects using the method of \citet{2006ApJ...638..622M, 2007ApJ...658...85M}. They found that type 2 quasars reside in halos up to 10 times as massive as those that host type 1 quasars.  However, \citet{2014MNRAS.442.3443D} showed that modifications to the angular mask (i.e., the description of the distribution of objects on the sky, with holes due to bright stars, bad/contaminated data, etc.)  applied to the \wise\ data in the same region significantly reduced this difference in clustering amplitude and thus halo mass.  The results of \citet{2014MNRAS.442.3443D} still showed that halo masses for type 2 objects are around three times greater when compared to type 1s.

Several effects make the clustering measurements of obscured quasars difficult to interpret.  The role of the angular mask has a two fold-effect on clustering measurements, because it is used to both properly weight the random catalog for normalization \citep[which can have a large effect on clustering results;][]{2011MNRAS.417.1350R, 2013MNRAS.435.1857L}, as well as remove regions of low quality data or artifacts \citep[which can also have a large effect on clustering results;][]{2014MNRAS.442.3443D}.  With CMB lensing cross-correlations, only the latter role of the mask is important, reducing the impact of subtle changes.  Additionally, the type 1 and 2 populations are not \textit{pure} because they are photometrically classified.  One significant source of contamination is from low redshift star-forming galaxies \citep[at a level of $\sim$10-15\%;][Hainline et al.\ 2014]{2007ApJ...671.1365H}.  There is also some confusion at low redshift between obscured and unobscured objects where the host galaxy's light becomes more important and obscured quasars resemble low-luminosity quasars that have a red component from host galaxy light.  This is where CMB lensing cross-correlations become a powerful tool --- not only does the measurement not depend heavily on the use of the quasars' angular mask to develop a random catalog for normalization of the clustering signal, but the shape of the lensing kernel also serves to down-weight the redshifts at which contamination and confusion are most significant.  As stars do not correlate with the CMB lensing, this method is also immune to stellar contamination.

\citet{2012PhRvD..86h3006S} showed the first significant detection of a cross-correlation between the CMB lensing convergence (from the Atacama Cosmology Telescope) and optically selected quasars from SDSS, over a relatively small 320 deg$^2$ region.  \citet{2013ApJ...776L..41G} used the 2500 deg$^2$ South Pole Telescope (SPT) CMB map to measure the cross-correlation with \wise\ selected quasars --- this was a \textit{mixed} type 1/2 sample, as no complementary optical data were available to split the samples.  The quasar bias measured in both cases was consistent with the results from clustering analyses.  In the latter study, the bias and halo masses of type 1 and 2 quasars are quite similar, as the mixed sample did not show a significantly higher bias compared to samples composed of only type 1 quasars.
 
In this work, we analyze the CMB lensing cross-correlation of a uniformly IR-selected quasar sample, using \wise\ and \planck.  Critically, we study the CMB lensing in a region where we can use optical imaging from the SDSS to split our quasar sample into obscured and unobscured sources, which has not been done previously.  This is a necessary and independent follow up to the recent angular autocorrelation measurements of \citet{2014ApJ...789...44D} and \citet{2014MNRAS.442.3443D}.  

We use a cosmology where $H_0 = 71$ km s$^{-1}$ Mpc$^{-1}$, $\Omega_{\textrm{M}}=0.27$, $\Omega_{\Lambda} =0.73$, $\Omega_{\textrm{b}} = 0.045$ and $\sigma_8 =0.8$ for all calculated parameters \citep{2011ApJS..192...18K}.  All magnitudes are given in the AB system unless otherwise specified.

\section{DATA}

\subsection{IR-Selected Quasars}

The quasar sample utilized in this study is the same as that used in \citet{2014MNRAS.442.3443D}.  We refer the reader there for full details of the sample and the angular mask applied to the data.  Here we briefly summarize the main points.

We start with all sources from the \wise\ all-sky catalog, which contains objects that have at least a 5$\sigma$ detection in one of the \wise\ bands ($W1$, $W2$, $W3$, and $W4$ at 3.4, 4.6, 12 and 22 $\mu$m, respectively), at least five observations, and are not flagged as spurious in at least one band.  The sources are limited to the range $135^{\circ} < \textrm{RA} < 226^{\circ}$ and $1^{\circ} < \textrm{DEC} < 54^{\circ}$, a region that overlaps SDSS optical imaging and is between the Galactic plane and the ecliptic pole (where the depth of \wise\ is less uniform due to more coverage).  We use the criteria of \citet{2012ApJ...753...30S} to select AGN candidates: $W1_{\textrm{AB}} - W2_{\textrm{AB}} > 0.16$ and $W2_{\textrm{AB}} < 18.32$ (corresponding to $W1_{\textrm{Vega}} - W2_{\textrm{Vega}} > 0.8$, $W2_{\textrm{Vega}} < 15.05$).  After applying the angular mask\footnote[2]{\textsc{mangle} polygon files marking the regions of data that have been removed can be found at \url{http://faraday.uwyo.edu/~admyers/wisemask2014/wisemask.html}}, we have a sample of 177,709 \wise-selected AGN over an area of 3,297 deg$^2$ (note that these numbers are slightly reduced by an additional cut in the next section)

To split this sample into type 1 and 2 sources, we match the \wise\ positions to the imaging of SDSS using a radius of 2$''$.  Using the extinction corrected $r$-band pipeline \textit{psfMags}, we split the type 1 and type 2 populations using a color cut at $r_{\textrm{AB}} - W2_{\textrm{AB}} = 3.1$ \citep{2007ApJ...671.1365H}.  Sources with no counterpart in SDSS are placed in the type 2 sample.  We performed tests to ensure these objects are real and not artifacts in \wise, the most convincing of which is that the angular clustering of the obscured sources is nearly identical (and well within the uncertainties) when these objects are included or excluded from the sample.  

We note that there are no strict definitions of ``obscured'' and ``unobscured'' quasars based purely on photometric data, as there are for spectroscopic samples.  \citet{2007ApJ...671.1365H} identified a clear bi-modality in the optical-IR colors of IR-selected quasars, and used this to define a split in the population.  This bi-modality is present in this sample as well \citep{2014MNRAS.442.3443D}.  While there is surely some overlap between the obscured and unobscured subsamples, it would only serve to reduce any differences in the derived bias and halo masses.  The samples are composed of 74,889 (42\%) type 2 and 102,740 (58\%) type 1 objects (again, these numbers are slightly reduced by an additional cut in the next section).

\begin{figure}
\centering
\hspace{0cm}
   \includegraphics[width=8cm]{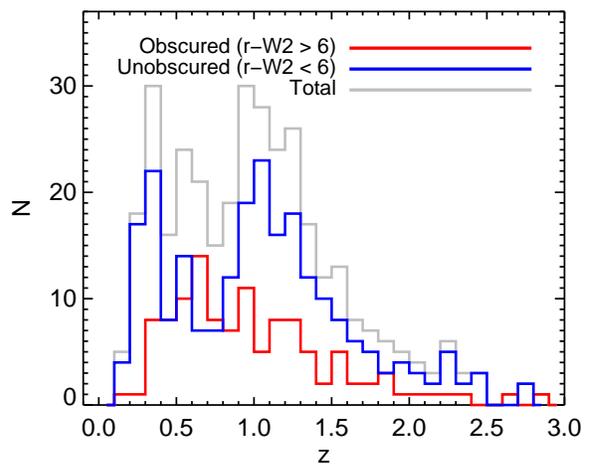}
   \vspace{0cm}
  \caption{The redshift distribution of the total, obscured and unobscured quasars, using identical \wise-selection criteria and a mask generated in the same way for objects in the Bo\"otes field.  Statistically, the obscured and unobscured $dN/dz$ is the same, allowing for accurate comparison of the two samples.\label{fig:z}}
\end{figure}

Because the quasar bias evolves with redshift, we must have samples of type 1 and 2 quasars with similar redshift distributions ($dN/dz$) in order to make fair comparisons.  Also, measuring the bias requires a model of the DM cross-correlation, which depends on $dN/dz$.  To estimate the redshift distributions of our samples we apply our selection criteria and mask to the Bo\"{o}tes survey field, which has extensive spectroscopy of AGN \citep{2012ApJS..200....8K}, as well as photometric redshift data from \spitzer\ IRAC \citep{2006ApJ...651..791B, 2011ApJ...731..117H}.  Detailed discussion of the redshift distributions are given in \citet{2014MNRAS.442.3443D}, and shown here in Figure~\ref{fig:z}.  An analysis of the effect of changes to the redshift distributions on our results is given in section 4.2.

\subsection{\planck\ CMB Maps}
\planck\ \citep{2011A&A...536A...1P} mapped the entire sky several times at nine frequencies, from 30 to 857 GHz, and released its first full data set in March 2013, with an update in December 2013.  \planck\ has produced the current state-of-the-art all-sky CMB map, with sensitivities down to $\mu$K levels and a beam size of $\sim 7'$.  This allows \planck\ to accurately measure gravitational lensing deflections, typically on scales of a few arcminutes.  

\begin{figure*}
\centering
\vspace{2cm}
   \includegraphics[angle=90,width=15cm]{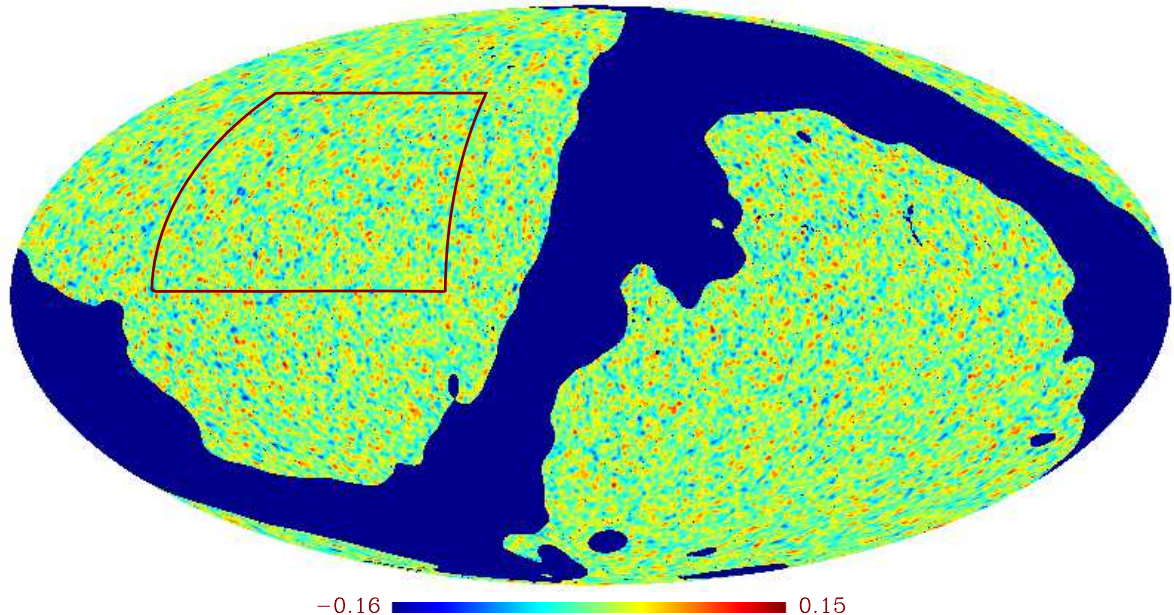}
   \vspace{0cm}
  \caption{The all-sky lensing convergence ($\kappa$) map from \planck, in equatorial coordinates.  The map has been smoothed with a 1$^{\circ}$ Gaussian for visualization, but the analysis is performed with the un-smoothed map.  The region used for the analysis here is shown by the black box.\label{fig:kappa}}
\end{figure*}

We start with one of the main data products of \planck, the all-sky lensing potential ($\psi$) map\footnote[3]{\url{http://irsa.ipac.caltech.edu/data/Planck/release_1/all-sky-maps/previews/COM_CompMap_Lensing_2048_R1.10/index.html}} \citep{2013arXiv1303.5077P}.  Using spherical harmonic transform tools in \textsc{HEALPix}\footnote[4]{\url{http://healpix.jpl.nasa.gov}}, this map is converted into a lensing convergence ($\kappa = \frac{1}{2} \nabla^2 \psi$) \ map, shown in Figure~\ref{fig:kappa} (all sky) and Figure~\ref{fig:map} (region used in this study).  The \planck\ maps are in \textsc{HEALPix} format, with $n_{side}=2048$, corresponding to a pixel size of $\sim$3 arcmin$^2$.

In order to estimate the errors on our lensing cross-correlation, we make use of the publicly available \planck\ simulated lensing maps\footnote[5]{Available at \url{http://irsa.ipac.caltech.edu/data/Planck/release_1/ancillary-data/HFI_Products.html\#hfisims}} \citep{2013arXiv1303.5077P}.  These maps contain 100 realizations of the lensing potential maps that accurately reflect the \planck\ noise characteristics.

\section{ANALYSIS \& RESULTS}

\subsection{Quasar Density}
\begin{figure*}
\centering
\hspace{0cm}
   \includegraphics[width=13cm]{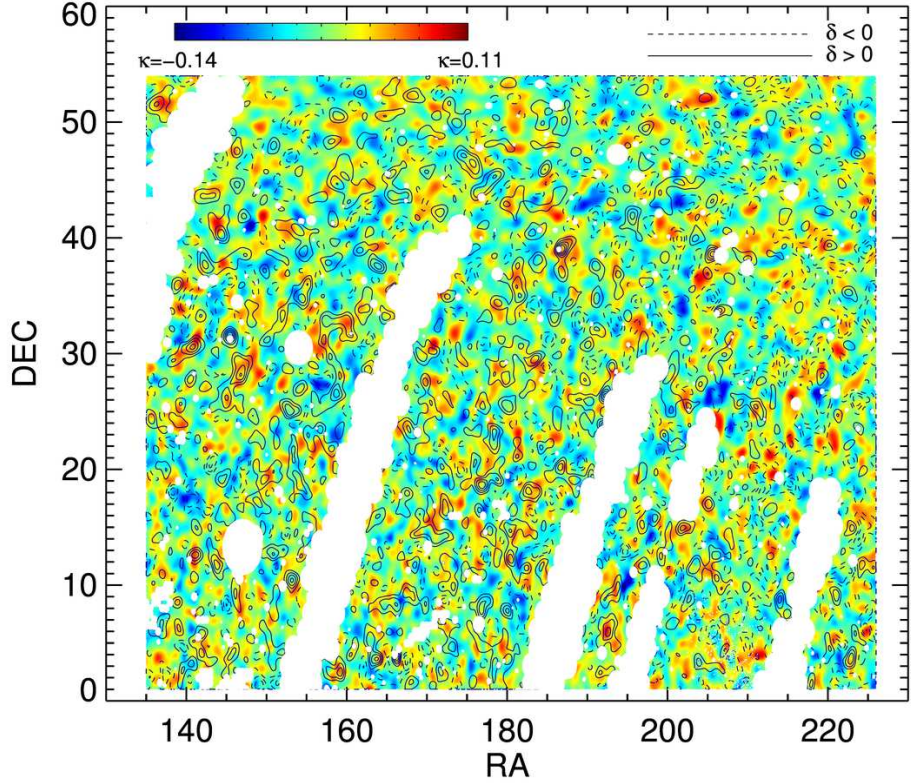}
   \vspace{0cm}
  \caption{The region used in this study \citep[see][]{2014MNRAS.442.3443D}.  The CMB lensing convergence $\kappa$ from the \planck\ lensing potential maps is shown as the colored contours, with blue indicating strong negative convergence and red indicating strong positive convergence (the convergence map has been smoothed with a $1^{\circ}$ Gaussian for visualization purposes).  Overlaid is a contour map of the fractional density of all \wise-selected quasars, with solid lines indicating over-densities and dashed lines indicating under-densities (also smoothed with a $1^{\circ}$ Gaussian for visualization).  The two are clearly correlated, as shown further in Figure~\ref{fig:binned}.\label{fig:map}}
\end{figure*}

Using the quasar sample described above, we generate a \textsc{HEALPix} density fluctuation map (with $n_{side} = 2048$, to match the \planck\ map resolution) of the total IR-selected sample, as well as the type 1 and 2 samples:
\begin{equation}
\delta = \frac{\rho -  \langle \rho \rangle}{\langle \rho \rangle}
\end{equation}
where $\rho$ is the density in a given pixel and $\langle \rho \rangle$ is the mean density in each case (53, 22, and 31 deg$^{-2}$ for total, obscured and unobscured samples, respectively).  The \planck\ mask (provided with the lensing map) is combined with the mask applied to the \wise\ data, only retaining the pixels that do not overlap any components of our mask.  While it is possible to estimate the area of fractional pixels where there is overlap, this step has its own systematic errors that can then carry through to the final bias measurement.  Since the area lost by removing these partial pixels is small (less than 200 deg$^2$), removing them is the simplest approach.  This step is analyzed further in Section 4.2.  Removing these partial pixels reduces the sample sizes to 169,945 (total), 71,535 (obscured), and 98,209 (unobscured), respectively.  The quasar density map (smoothed with a 1$^{\circ}$ Gaussian for visualization purposes) in the region of interest is shown as the solid ($\delta > 0$) and dashed ($\delta < 0$) lines in Figure~\ref{fig:map}, overlaid on the map of $\kappa$ in this region.  

In fields of finite size, various effects can impact the mean density calculations and thus the cross-correlations.  For example, there are known corrections to the mean density calculated in small areas \citep[the integral constraint;][]{1977ApJ...217..385G}.  Given that the integral constraint is inversely proportional to the square of the area, and our region is over 3000 deg$^2$, this correction is at a sub-percent level.  Other systematics, such as depth variations and improper masking of the data, can also play a role.  However, this region was chosen in \citet{2014ApJ...789...44D} precisely because it has an even depth in \wise\, and is far from the stellar contamination of the Galactic plane.  Combined with the additional masking of \citet{2014MNRAS.442.3443D}, this sample is likely one of the cleanest and uniform of its kind, and systematic errors in the density calculations should be minimal.

\subsection{CMB Lensing Cross-Correlation}
\begin{figure}
\centering
   \includegraphics[width=7.5cm]{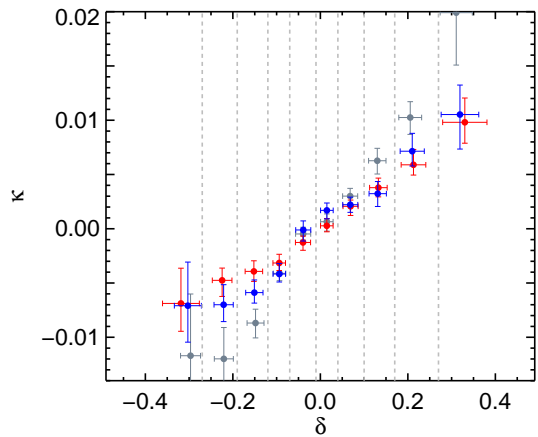}
  \caption{Binned values of the ($1^{\circ}$ Gaussian smoothed) quasar fractional density for all \wise-selected quasars (gray points), the obscured sample (red points), and the unobscured sample (blue points), versus the average ($1^{\circ}$ Gaussian smoothed) lensing convergence in the corresponding pixels.  Bins in $\delta$ are marked with gray lines, and correspond to the bins used in Figure 4 of \citet{2013ApJ...776L..41G}.  The points are placed at the average value of $\delta$ in each bin, and horizontal error bars indicate the scatter in $\delta$ within each bin. Vertical error bars are from repeating the calculation with each of the 100 simulated noise maps (each also smoothed with a $1^{\circ}$ Guassian) in place of the actual convergence map.\label{fig:binned}}
\end{figure}

Quasars trace peaks in the matter density field and should be denser in regions of enhanced lensing convergence.  Qualitatively, it is immediately clear in Figure~\ref{fig:map} that this is the case.  This is shown even more clearly in Figure~\ref{fig:binned}, in which we have smoothed both the density and convergence maps with a $1^{\circ}$ Gaussian, and found the pixels in which the values of $\delta$ fall in the bins marked with vertical gray lines.  We take the average value of $\delta$ in each bin, and the average value of $\kappa$ in the corresponding pixels, and plot these against each other --- $\kappa$ and $\delta$ are strongly correlated for all of the samples.  Note that because the \planck\ maps are noise-dominated, these correlations are only visually apparent when the maps are smoothed and/or stacked.

To quantify this correlation, we follow the formalism of measuring the CMB lensing cross-power spectrum presented fully in \citet{2012ApJ...753L...9B} and \citet{2012PhRvD..86h3006S}, and reviewed in \citet{2013ApJ...776L..41G} --- we include a summary of the method here.

The lensing convergence ($\kappa$) in comoving coordinates ($\chi$) along a line of sight $\mathbf{\hat{n}}$ can be expressed as an integral over the fractional over-density of matter ($\delta(\mathbf{r},z)$) multiplied by the lensing kernel ($W^{\kappa}$):

\begin{equation} \kappa(\hat{\mathbf n}) = \int d\chi
W^\kappa(\chi)\delta(\chi{\hat{\mathbf n}},z(\chi)). 
\end{equation}

\noindent The lensing kernel is \citep{2000ApJ...534..533C, 2003ApJ...590..664S}:

\begin{equation} W^\kappa(\chi) = \frac{3}{2}\Omega_{\rm
m}\left(\frac{H_0}{c}\right)^2\frac{\chi}{a(\chi)}\frac{\chi_{\rm CMB} - \chi}{\chi_{\rm CMB}},
\end{equation}
\noindent where $a(\chi) = (1+z(\chi))^{-1}$ is the scale factor, and $\chi_{\rm CMB}$ is the co-moving distance to the CMB (13.98 Gpc in our cosmology).  Fluctuations in the quasar density can be expressed as:

\begin{equation} q(\hat{\mathbf n}) = \int d\chi
W^q(\chi)\delta(\chi{\hat{\mathbf n}},z(\chi)), 
\end{equation}

\noindent where $W^q(\chi)$ is the distribution kernel of quasar hosts:

\begin{equation} W^q(\chi) = \frac{dz}{d\chi}\frac{dN(z)}{dz} b_q(\chi).
\end{equation}

\begin{figure}
\centering
   \includegraphics[width=8cm]{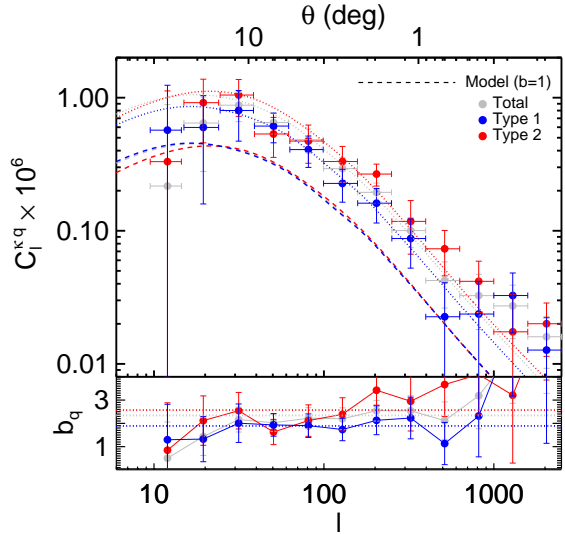}
  \caption{The top panel shows the cross-correlation results for the total, type 2, and type 1 samples.  The dashed lines show the model DM cross-correlation (Equation 6) for each sample, using the corresponding $dN/dz$.  The dotted lines show the fits for each sample, used to measure the bias.  The bottom panel shows the data divided by the model, another way of visualizing the bias.  The horizontal lines mark the adopted bias values shown in Table~\ref{table:results}.\label{fig:crosscorr}}
\end{figure}

\noindent Here, $dN/dz$ is the normalized redshift distribution of the quasar population, which has bias $b_q$. The cross-power at a Fourier mode $l$ is

\begin{equation} C^{\kappa q}_l = \int dz
\frac{d\chi}{dz}\frac{1}{\chi^2}W^\kappa(\chi)W^q(\chi)P\left(\frac{l}{\chi},z
\right) \end{equation}

\noindent where $P(k=l/\chi,z)$ is the matter power spectrum (e.g.\ Eisenstein \& Hu 1999) --- we use the non-linear matter power spectrum from CAMB\footnote[6]{\textit{Code for Anisotropies in the Microwave Background} (\url{http://lambda.gsfc.nasa.gov/toolbox/tb_camb_ov.cfm})} \citep{2000ApJ...538..473L}.  Equation 6 gives us the model cross-power spectrum for the underlying distribution of DM (with a bias of 1), shown as the dashed lines in Figure~\ref{fig:crosscorr} for each $dN/dz$.

The cross-power $C_l^{\kappa q}$ is measured for the data by taking the Fourier transform of both the lensing convergence map ($M_{\kappa}$) and the fractional density map ($M_q$) and multiplying them\footnote[7]{Using tools in the \textsc{HEALPix} package.}:

\begin{equation} C_l^{\kappa q} = \left< \mathrm{Re}( \mathcal{F}({\sf
M_\kappa})\mathcal{F^*}({\sf M_q}))|_{\mathbf{l}\in l} \right>
\end{equation}

\noindent where $\mathbf{l}\in l$ describes the binning.  We present our results with 5 bins in $l$ per dex, with the edge of the first bin starting at $l=10$.  Uncertainties in $C^{\kappa q}_{l}$ are derived by repeating the calculation using the 100 simulated noise maps from \planck\ (see section 4.2 for an analysis of possible additional systematic errors).  We derive covariance matrices ($\mathbb{C}(l_i,l_j) = \mathbb{C}_{ij}$):
\begin{equation}
\mathbb{C}_{ij} = \frac{1}{1-N} \left[ \sum_{k=1}^{N} (C_{l_i,k}^{\kappa q} - C_{l_i}^{\kappa q}) (C_{l_j,k}^{\kappa q} - C_{l_j}^{\kappa q}) \right],
\end{equation}
where $N$ is the number of simulation cross-correlations and $C_{l,k}^{\kappa q} - C_{l}^{\kappa q}$ is the cross-correlation from each noise simulation.  We adopt the square root of the diagonal elements of $\mathbb{C}_{ij}$ as the 1$\sigma$ errors on the cross-correlation.  The cross-correlation results are shown for all samples in Figure~\ref{fig:crosscorr}.

\begin{figure}
\centering
   \includegraphics[width=8cm]{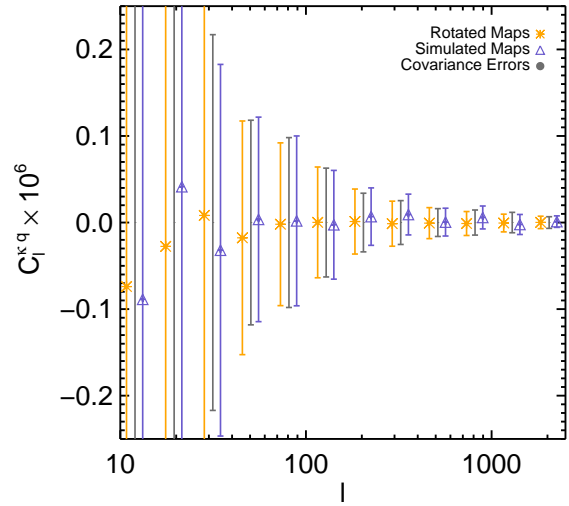}
  \caption{A comparison of various methods of estimating the errors in $C_l^{\kappa q}$ (using the total \wise-selected sample) --- the simulated noise maps from \planck\ and rotating the real convergence map.  Points are placed at the mean value from each simulation or rotation, and error bars indicate the 1$\sigma$ scatter.  Points are offset slightly in $l$ for clarity.  The gray bars indicate the size of the adopted error bars from the diagonal of the covariance matrix.  The cross-correlation is null for all $l$ in both cases, and the error bars are consistent.\label{fig:errs}}
\end{figure}

As an additional check that the detection of the cross-correlation is real and the errors are reasonable, we also measure the cross-correlation with multiple rotations of the \planck\ convergence map.  This preserves any systematic errors in the map, in addition to the random instrument errors.  We utilize a total of 34 rotations, in 20$^{\circ}$ increments in Galactic longitude (17 rotations), and again in 20$^{\circ}$ increments in Galactic longitude with an additional 180$^{\circ}$ rotation in Galactic latitude (17 rotations).  We find that the mean cross-correlation from the rotations is null on all scales, and the standard deviation in each bin is nearly identical to the 1$\sigma$ errors from the covariance matrix.  The consistency between the different methods is shown in Figure~\ref{fig:errs}.

\subsection{Bias and Halo Masses}
For a given cosmology and a well-constrained $dN/dz$, Equation 6 depends only on the quasar bias $b_q$ (see section 4.2 for analysis of the effects of uncertainties in the cosmology and $dN/dz$).  We fit Equation 6 to the data using a $\chi^2$ minimization and the full covariance matrix:
\begin{equation}
\chi^2 = \sum_{i.j} (C_{l_i}^{\kappa q} - C_{l_i, \textrm{model}}^{\kappa q}) \mathbb{C}_{ij}^{-1} (C_{l_j}^{\kappa q} - C_{l_j, \textrm{model}}^{\kappa q}).
\end{equation}

\begin{figure*}
\centering
   \includegraphics[width=14cm]{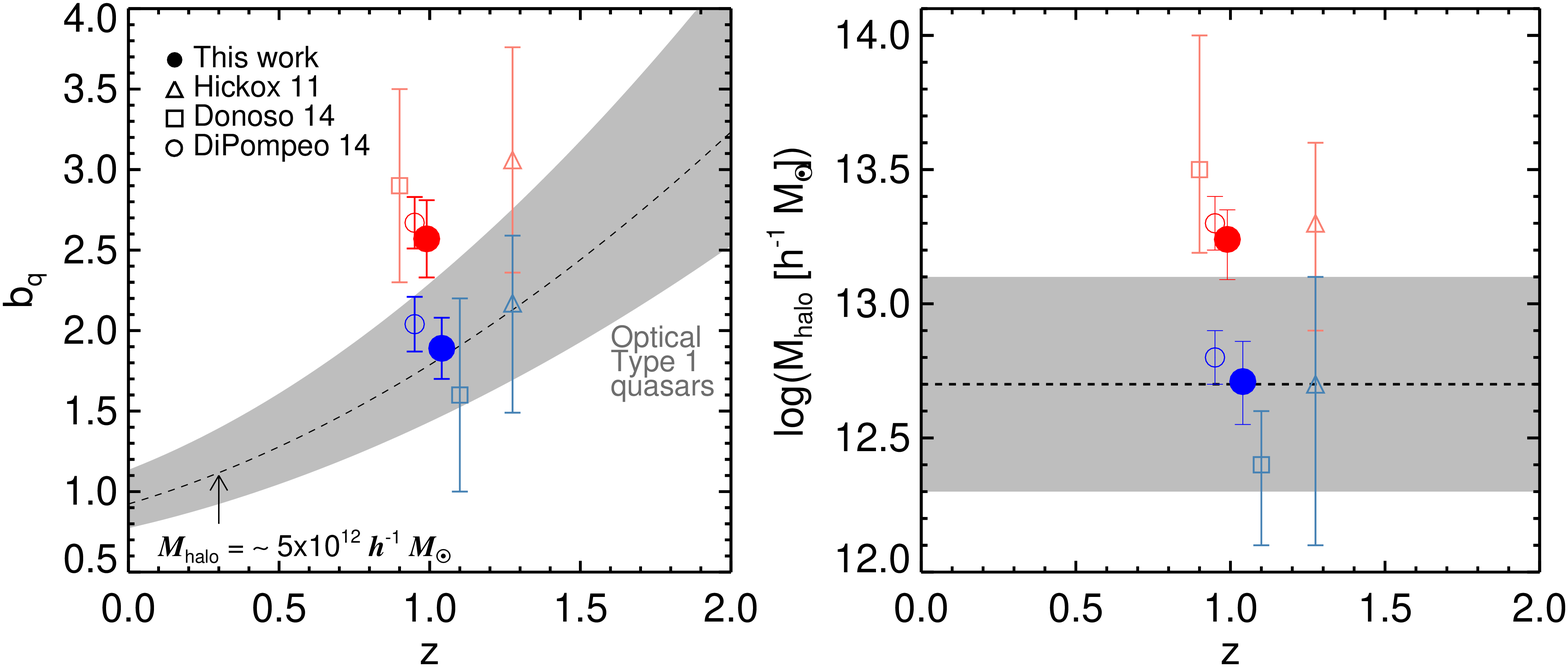}
  \caption{A comparison of the quasar bias (left) and corresponding halo mass (right) found in this work with recent angular clustering results. \citet{2014ApJ...789...44D} and \citet{2014MNRAS.442.3443D} analyze the same region as we do here --- \citet{2014MNRAS.442.3443D} analyzed the same sample \citep[the][results are shifted slightly in $z$ for clarity]{2014MNRAS.442.3443D}.  The results from this sample using clustering and CMB lensing are nearly identical, with a very slight reduction in bias and halo mass for both samples.\label{fig:resultcompare}}
\end{figure*}

Fits are performed over the range $10 < l < 2000$ --- changing this region shifts the results slightly, but always within the errors.  As there is only one free parameter, we derive 1$\sigma$ errors on $b_q$ by the range for which $\Delta \chi^2 =1$.  The results are listed in Table~\ref{table:results}, and shown along with other recent measurements in the left panel of Figure~\ref{fig:resultcompare}.  An analysis of various model assumptions and systematics that can impact the bias measurements is given in section 4.2.

We follow the same procedure as \citet{2014MNRAS.442.3443D} and \citet{2007ApJ...658...85M} to convert the bias to an average halo mass ($M_h$).  This method uses the ellipsoidal collapse model of \citet{2001MNRAS.323....1S}\footnote[8]{We choose this model in order for more direct comparison with previous work by other authors.  However, we point out that we are generally looking for differences in halo masses, which is less sensitive to our model choice.  Using the \citet{2005ApJ...631...41T} model changes the halo masses by $\sim$0.1 dex.}, and models the linear power spectrum including the effects of baryons from \citet{1998ApJ...496..605E}.  The masses are calculated both at the mean redshift for each sample, which is very nearly $z \sim 1$ in all cases, as well as at the effective redshift considering the weight of the lensing kernel.  The true effective redshift is likely somewhere between these values (see section 4.2 and Figure~\ref{fig:effectivez}).  The derived halo masses are listed in Table~\ref{table:results} and shown in the right panel of Figure~\ref{fig:resultcompare}.

\begin{table}
\centering
  \caption{Quasar bias and dark matter halo mass.}
  \label{table:results}
  \begin{tabular}{lccc}
  \hline
  Sample                      &  $\langle z_{\textrm{eff}} \rangle$ &      $b_q$             &  $\log (M_{h}/M_{\odot}$ $h^{-1}$)                  \\ 
\hline                           
    \vspace{0.15cm}
   Total                          &                 1.02                     & 2.33 $\pm$ 0.15 & $13.07^{+0.11}_{-0.07}$   \\ 
   \vspace{0.15cm}
   Obscured                 &                  0.99                    & 2.57 $\pm$ 0.24 & $13.24^{+0.15}_{-0.13}$    \\
   \vspace{0.15cm}
   Unobscured            &                  1.04                    & 1.89 $\pm$ 0.19 & $12.71^{+0.14}_{-0.15}$    \\                         
\hline
   \vspace{0.15cm}
   Total                          &                  1.50                    & 2.33 $\pm$ 0.15 & $12.59^{+0.10}_{-0.09}$   \\ 
   \vspace{0.15cm}
   Obscured                 &                  1.47                    & 2.57 $\pm$ 0.24 & $12.77^{+0.17}_{-0.12}$    \\
   \vspace{0.15cm}
   Unobscured            &                  1.51                    & 1.89 $\pm$ 0.19 & $12.21^{+0.15}_{-0.18}$    \\                         
\hline
   \end{tabular} 
    \\
{
\raggedright    
The top half of the table shows the conversion from bias to halo mass using the the mean redshift of each sample, and the bottom half shows the conversion using the mean redshift weighted by the lensing kernel (see section 4.2 and Figure~\ref{fig:effectivez}).  \\
 }
\end{table}

\section{DISCUSSION}
\subsection{Comparison with previous results}

Our CMB lensing cross-correlation results are nearly identical to those using angular clustering, and agree very well with previous measurements from other samples (Figure~\ref{fig:resultcompare}).  We confirm that type 2 objects have a higher bias, and given the reasonable match in redshift distributions, higher halo masses.  As in \citet{2014MNRAS.442.3443D}, we find that type 2 objects have halo masses a factor of $\sim$3 times greater than the halo masses of type 1 objects, a significant reduction in the factor of $\sim$10 found by \citet{2014ApJ...789...44D}.

\citet{2013ApJ...776L..41G} measured the bias for a sample of similarly \wise-selected type 1 \& 2 quasars in the SPT field, finding  $b_q = 1.61 \pm 0.22$ assuming no evolution over the redshift range of the sample (as is done here; see section 4.2), with consistency between cross-correlations with the SPT and Planck lensing maps. This is lower than the $b_q = 2.33 \pm 0.15$ in this work; if we perform our analysis in the SPT region, applying the same mask to the WISE data used in this work, we measure $b_q = 2.31 \pm 0.22$, in agreement with what we find in the larger region studied here. A possible explanation for the higher bias we measure for \wise-selected quasars is the stricter masking we employ.

As discovered in \citet{2014MNRAS.442.3443D}, it is necessary to discard objects in regions surrounding artifacts/contamination in the \wise\ data when using \wise\ to build quasar samples, as it seems that the \wise\ flags may be too conservative. With our full mask of the \wise\ data, the area of the SPT field drops from 2500 deg$^2$ to $\sim$2000 deg$^2$. This is a difference of $\sim$20\%, and could account for some of the apparent inconsistency. If objects in these regions are indeed artifacts, they should be completely uncorrelated with the CMB and their inclusion should reduce the strength of the cross-correlation. As a test, we do not fully mask the regions around \wise\ flagged data, only discarding the flagged data themselves \citep[the ``partial'' mask described in][]{2014MNRAS.442.3443D} in the SPT field. Repeating the cross-correlation measurement, this drops the bias in the SPT field to $b_q = 2.1 \pm 0.2$. If we go another step further, and do not remove any regions of large galactic extinction, low \wise\ median coverage, or Moon contamination, the bias drops to $b_q = 2.0 \pm 0.3$. It is likely that the combination of contamination and effective area explains the difference in results between our measurement and those of \citet{2013ApJ...776L..41G}.

\subsection{Caveats and additional sources of error}
There is another possible interpretation of the difference in measured bias rather than a difference in typical halo mass --- there may be a difference in the shape of the halo occupation distribution (HOD) for obscured and unobscured quasars, or there may be an observational effect that causes us to sample different ranges of the HOD in the two samples.  It seems that a log-normal HOD is an adequate description for unobscured quasars, and that there is no clear dependence on other properties such as luminosity or redshift \citep[e.g.][]{2012MNRAS.424..933W, 2005ApJ...633..791Z, 2007ApJ...659....1Z, 2013ApJ...779..147C}.  While there is no observational evidence to suggest that the HOD of obscured sources behaves differently, if it does then this could cause the bias to appear different while the average or typical halo mass is the same.  Theoretically, this possibility might be tested via careful modeling of the HOD of obscured quasars \citep[as has been conducted for AGN and for black holes in general, e.g.][]{2011MNRAS.416.1591D, 2012MNRAS.419.2657C}. Observationally, progress is likely to only be made with full redshift information for individual sources in the obscured sample. Obtaining redshifts for obscured quasars
in a wide-area survey is likely to prove taxing, but it would allow further constraints on the HOD of obscured sources through higher-order correlation functions or direct measurements of the Mean Occupation Function of obscured quasars \citep[c.f.][]{2013ApJ...779..147C}. Such measurements are far more sensitive to the shape and range of the HOD, particularly at the highest halo masses \citep[e.g.][]{2002ApJ...575..587B, 2004ApJ...614..527Z}.

Another important consideration for interpreting our results is errors in the redshift distributions.  There are several factors that can bias $dN/dz$, particularly for the type 2 sources.  For example, the inclusion of objects that are undetected by SDSS may shift the actual $dN/dz$ to a higher average $z$.  Based on analysis of the objects detected in the Bo\"{o}tes field that fall below the SDSS $r$-band completeness limit, we argued in \citet{2014MNRAS.442.3443D} that this will at most shift the mean redshift by $\sim$ 0.1.  How much does this affect the bias measured by the CMB lensing cross-correlation?  

\begin{figure}
\centering
   \includegraphics[width=6cm]{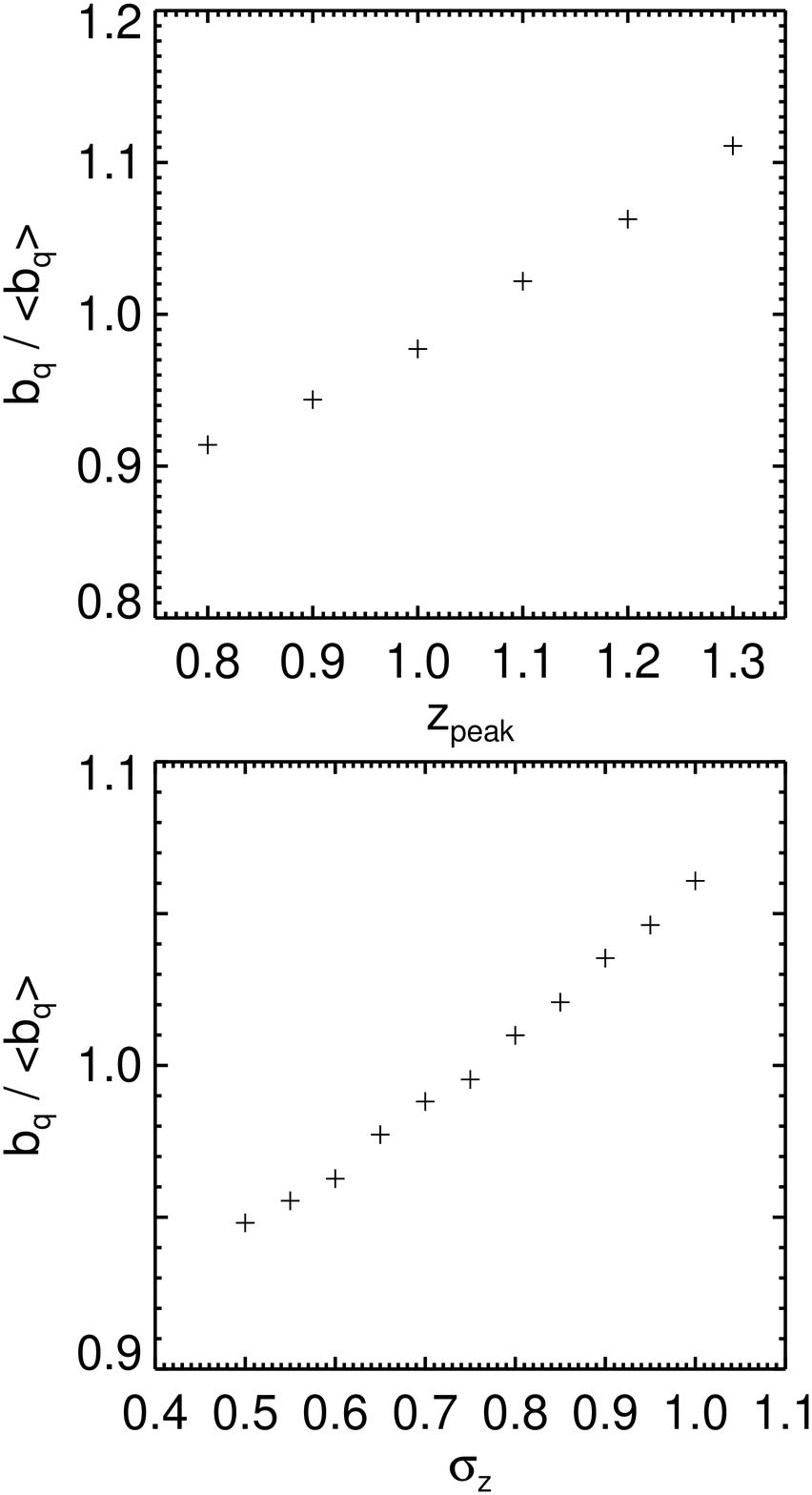}
  \caption{The effects of changing the mean $z$ (top) or width of $dN/dz$ on the measured bias.  Changes are shown relative to the mean over all simulated $z$ distributions.\label{fig:biaschange}}
\end{figure}

To quantify this, we generate mock Gaussian distributions of $z$ with various parameters, treat them as we do the real redshift distributions when generating our model DM cross-correlation, and re-fit the bias to the total \wise-selected sample.  The mock $z$ distributions range from $0.8 < \langle z \rangle < 1.3$ (in steps of 0.1 while holding $\sigma_z = 0.55$, roughly the value of the actual $z$ distribution), or in width from $0.5 < \sigma_z < 1.0$ (in steps of 0.05, holding $\langle z \rangle = 1$).  The effect of these changes is shown in Figure~\ref{fig:biaschange}. The mean redshift has a larger effect than the width of the distribution.  However, unless the inferred redshift distributions for our sample are incorrect by a large amount, the effect on the measured bias is at most a few percent.  While this would serve to reduce the magnitude of the difference between the bias of type 1 and 2 objects, it is unlikely to explain it completely.

\begin{figure}
\centering
   \includegraphics[width=8cm]{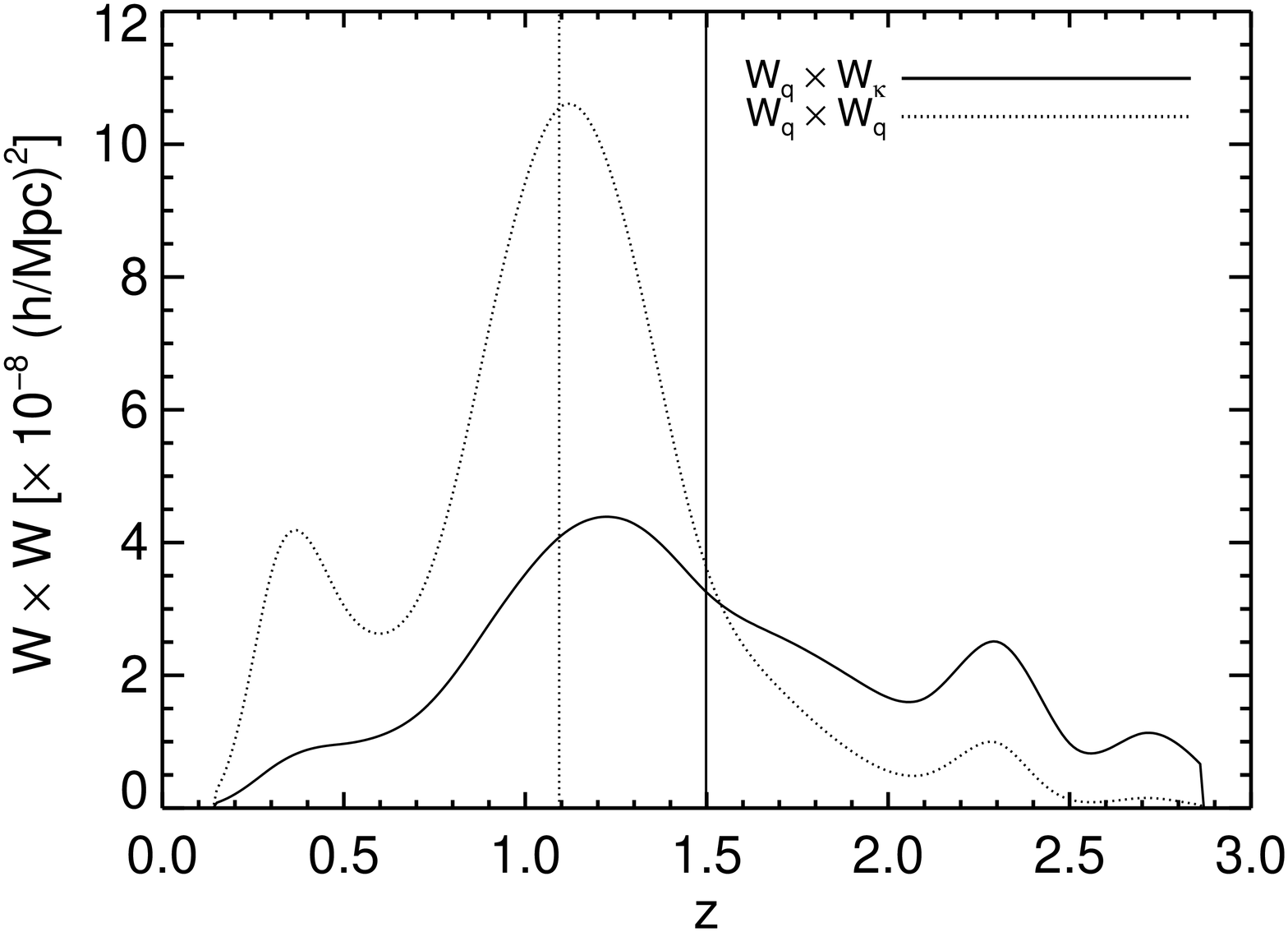}
  \caption{The effective redshift as weighted by the quasar distribution kernel squared ($W_q^2$; dotted line) and a multiplication of the lensing kernel and the quasar distribution kernel ($W_{\kappa}W_q$; solid line), under the assumption that each redshift contributes to the clustering or cross-correlation signal equally.  For quasar autocorrelation studies, the effective redshift is the same as the mean redshift of the sample (vertical dashed line at $z=1.09$), but for lensing cross-correlations it is pushed to larger values because the lensing kernel peaks near $z \sim 2$ (vertical solid line at $z=1.5$).\label{fig:effectivez}}
\end{figure}

Not only can errors in the redshift distribution affect our results, but the conversion of the bias into a halo mass also depends on the adopted effective redshift.  In the simplest case, this is just the mean redshift of the sample, which we have used in our analysis.  However, while this is the obvious choice when analyzing the quasar autocorrelation function (where the effective redshift depends on just the quasar distribution kernel as $W_q^2$), the redshift weighting for CMB lensing cross-correlations depends on $W_{\kappa} W_q$ (see equation 6).  Because the weight of the lensing kernel ($W_{\kappa}$) peaks near $z \sim 2$, this will increase the effective $z$.  This effect is illustrated in Figure~\ref{fig:effectivez} using $dN/dz$ for the total \wise-selected sample.  The effective redshift changes from $z \sim 1$ to $z \sim 1.5$.  However, this also assumes that all redshifts contribute to the cross-correlation signal equally.  Without full redshift information to analyze the cross-correlation in bins of redshift, we cannot retrieve the \textit{true} effective redshift, and so we prefer the value with the fewest assumptions --- the mean redshifts of the samples.  However, for completeness we include in Table~\ref{table:results} halo mass conversions using the larger effective redshift considering the weight of the lensing kernel.

As noted in section 3.1, we only include \textsc{HEALPix} pixels that are unaffected by the mask applied to the \wise\ data in our analysis.  As a way to quantify the effects of changing the overall mask applied to the data, we repeat the cross-correlation and bias measurements including the partial pixels that do overlap the \wise\ mask.  If we track the area of partial pixels (again, a step which has its own errors associated with it), and include these pixels in our measurements, we find that the bias can vary by roughly 9\% despite the fact that only $\sim$4\% of pixels are affected.  This highlights the importance of subtle changes in the mask in bias measurements.

Finally, the analysis thus far has assumed a fixed cosmology, when in reality the cosmological parameters used to generate the model power spectra and thus calculate the bias have their own uncertainties.  We note however, that changes in cosmology will affect all of the samples in a similar way, and thus will not change our overall conclusions regarding the difference in halo mass between obscured and unobscured quasars.  In terms of the absolute bias, if for example we vary $H_0$ by $\pm2$, slightly larger than the uncertainty in \citet{2011ApJS..192...18K}, we find that the bias changes by less than 10\%, remaining roughly within the current error bars.

\subsection{Quasar lifetimes}

Comparing the number density of a quasar population to the number density of halos at the typical halo mass can constrain the length of the quasar duty cycle \citep[abundance matching, e.g.][]{1999ApJ...523...32C, 1999ApJ...520..437K, 2004MNRAS.353..189V, 2006ApJ...643...14S, 2010MNRAS.404.1111G}.  The median bolometric luminosity of IR-selected quasars at this flux limit is $L_{\textrm{bol}} \sim 10^{46}$ erg s$^{-1}$ \citep[][Hainline et al.\ 2014, in press]{2011ApJ...731..117H}.  Using the bolometric luminosity function of \citet{2007ApJ...654..731H} at $z \sim 1$, this gives a space density of \wise-selected quasars of $2 \times 10^{-5}$ Mpc$^{-3}$, of which $\sim$60\% are unobscured and $\sim$40\% are obscured.  For the typical halo masses determined above and the mass function of \citet[][]{2001MNRAS.323....1S}, we find space densities of $dn/d \log(M) = (4.6^{+2.2}_{-1.4}) \times 10^{-4}$ Mpc$^{-3}$ (unobscured) and $(1.0^{+0.5}_{-0.4}) \times10^{-4}$ Mpc$^{-3}$ (obscured).  Since the majority of our sample is in the range $0.5 < z < 1.5$, which spans about 4 Gyr of cosmic time, these abundances imply lifetimes of $123^{+ 56}_{- 39}$ Myr for the unobscured phase and $302^{+191}_{- 99}$ Myr for the obscured phase.  

These results are consistent with previous quasar lifetime results, including \citet{2014MNRAS.442.3443D}.  We confirm that the obscured phase makes up a significant portion of the quasar duty cycle, and may in fact last more than twice as long as the unobscured phase.  This can provide constraints on models in which quasars evolve from obscured to unobscured.

\section{CONCLUSIONS}

It has now been shown using two independent methods that there is indeed a difference in bias and typical halo mass between type 2 (obscured) and type 1 (unobscured) quasars at $z \sim 1$.  Careful removal of regions of poor \wise\ data is needed to build clean quasar samples to accurately quantify this difference, as both CMB-lensing cross correlations and angular clustering measurements have illustrated.  On average, type 2 objects are found in halos that are $\sim$3 times more massive than those of type 1 objects at $z \sim 1$.  This suggests that the obscured phase is longer than the unobscured phase in quasars, with both still on the order of $\sim$1\% of the Hubble time.  

Given that \wise\ and \planck\ both cover the whole sky, the next natural step is to provide the tightest possible constraints on the IR-selected quasar bias using as much area as possible.  In areas where optical coverage also exists --- over the rest of the SDSS and the Dark Energy Survey field, for example --- we can further constrain the bias of obscured and unobscured quasars.  As CMB lensing cross-correlations are the most robust, direct way to measure the quasar bias, these measurements will provide the best constraints on models of quasar and galaxy evolution.  Additional work by our group will also further constrain the $dN/dz$ of \wise-selected quasars (Hainline et al.\ 2014, in press), as well as improve photometric redshift estimation for those with optical counterparts, allowing us to make the first measurements of the CMB-lensing cross correlation as a function of redshift and study the cosmic evolution of the IR-selected quasar bias.

\section*{Acknowledgements}
MAD, ADM, RCH, KNH, and SWH were partially supported by NASA through ADAP award NNX12AE38G and by the National Science Foundation through grant numbers 1211096 and 1211112. MAD, ADM and SWH were partially supported by EPSCoR award NNX11AM18A.  JEG acknowledges support from the Royal Society.

\bibliography{/Users/Mike/Research/bibliographies/full_library.bib}

\label{lastpage}

\end{document}